\documentstyle[12pt,epsf]{article}
\input epsf
\begin{document}

\begin{titlepage}
%\centerline{\large\bf INSTITUTE OF THEORETICAL AND EXPERIMENTAL PHYSICS}
%\vspace{1cm}
\hspace{102mm}{\bf ITEP N 58-95}
\vspace{2cm}

\centerline{\large\bf H.B.Nielsen}
\centerline{\large\bf Niels Bohr Institute, Copenhagen, Denmark}

\vspace{3mm}

\centerline{\large\bf A.V.Novikov, V.A.Novikov, M.I.Vysotsky}
\centerline{\large\bf ITEP, Moscow,Russia}

\vspace{2cm}

\centerline{\large\bf HIGGS POTENTIAL BOUNDS ON EXTRA}
\centerline{\large\bf QUARK-LEPTON GENERATIONS}
\vspace{2cm}
\begin{abstract}
%\vspace{5mm}

 We consider the bounds for  the values of higgs mass $M_H$ and of
 the mass of the extra quarks and leptons $M_{extra}$ derived from
 the stability of vacuum and from the absence of Landau pole in Higgs
 potential. We find that in the case of the absence of new physics up
 to the GUT scale the bounds for the mass of the 4th generation are
 so restrictive that the negative result of CDF search for extra quarks
 closes the window for fourth generation.  In
 the case of the absence of new physics up to $10^5$ GeV we get weaker but
 still nontrivial bounds on $M_H$ and $M_{extra}$ as well.
\end{abstract}
% \newpage
\vspace{2cm}

\centerline{\large\bf Moscow - 1995}
%\vspace{1cm}
% $~~$

\end{titlepage}

\newpage
% $~~$
%\vspace{60mm}

 With the discovery of the top quark at FNAL collider \cite{1} the
 third fermion generation is completed now and the question naturally
  arises: whether
  there exist more generations in nature (fourth, fifth and
  so on). Below we suppose that extra generations do exist.
  The severe bound on extra neutrinos obtained at LEP1 demonstrates
  that a new neutral lepton should be heavy, $M_N>M_Z/2 = 45$ GeV.
  Another bound on the spectrum of new fermions come from precision
  measurements of $M_W$ and $Z$-boson properties. Since the value of
  $M_{top}$ from direct measurements nicely coincides with that
  obtained from analyzing precision data \cite{2}, then the  masses of
  weak isopartners ($T$ and $B$, $E$ and $N$) should be  almost
  degenerate.  In the opposite case their contributions to
  loops will be enhanced by the factor $\sim M^2_T - M^2_B$
  and will destroy a successful description of
  the precision data by the electroweak radiatively corrected
  formulas.
  On the other hand, the existence of the fourth
  generation with degenerate masses of quarks and leptons do not
  disturb a successful description of the data even for a very light
  extra generation with $M_4 = 50$ GeV \cite{3}. The absence of
  decoupling in the electroweak theory leads to a finite contribution
  to $Z$-boson parameters even for an infinitely massive fourth
  generation, however experimental accuracy is not enough to exclude
  the existence of extra generations.
  Taking into account this comment and in order to simplify the problem by
maximal
  diminishing the number of parameters  we will consider  the
  fourth generation of leptons and quarks with the degenerate mass
  $M_4$.  When we discuss more than one extra generation
  we will assume that all masses of new generations are degenerate.
  From LEP1 data we know, that $M_4>M_Z/2 = 45$ GeV. If new quarks
  are substantially mixed with the known light quarks then their mass
  can be effectively bounded by CDF and D0 searches for top quark
  since decays $B \to cW,\; T \to bW$ are analogous to $t \to bW$
  decay.  So we can estimate that the mass of such a new generation
  should be larger than, say, 150 GeV.
  The other possibility is that
  the fourth generation quarks are not mixed with the known quarks.
  To avoid the bounds for existence on the Earth of the absolutely
  stable new particles produced at the time of Big Bang we should
  propose that the new quarks are mixed with the light ones.
  However,
  the mixing angles are so small that the new particles leave FNAL
  detectors without decays.
   These, so to say, "decoupled" extra generation
  particles should be heavier than 45 GeV in order to satisfy LEP
  direct search bounds. In a dedicated analysis
  \cite{10} CDF ruled out the
    existence of stable, pair-produced colour triplets with masses
  $50 {\rm GeV} < m < 139$ GeV at 95\% c.l. Our results demonstrate
  that under this condition the existence of new generation will
  inevitably lead to appearence of new physics at a scale below
  $10^{10}$ GeV.

  Armed with the knowledge of experimental lower bounds on the masses of new
  fermions we will now get theoretical upper bounds on their
  masses.  The point is that such  heavy fermions will greatly deform
  the Higgs boson potential through radiative corrections
  \cite{300},\cite{301}. With the account of radiative corrections
  the renormalization group improved Higgs boson potential is usually
  presented in such a form \cite{301}:
  \begin{equation}
  V(\Phi) =
     -\frac{1}{2}\mu^2(t) G^2(t) \Phi^2 + \frac{1}{4}\lambda(t)
  G^4(t) \Phi^4\;\;, \label{1} \end{equation} where $t =
\ln(\Phi/\eta)\;,\; \eta = 246$ GeV and $G(t)$ is determined by
anomalous dimension of field $\Phi$.
  For $\Phi \sim \eta$ the initial
values of $\mu$ and $\lambda$ govern $V(\Phi)$ behavior while for $\Phi \gg
\eta$ the radiative corrections become essential.  Higgs boson mass  is
defined as a second derivative of the potential at minimum. Crucial for the
selfconsistency of the theory is the $\lambda(t)$ behavior. If $\lambda(t)$
becomes negative at some value $\Phi_0$ -- then the $V(\Phi)$ minimum at
$\Phi = \eta$ becomes unstable and the experimental value of Fermi coupling
constant $G_F$ cannot be obtained.  On the other hand, to avoid strong
interactions in Higgs sector we demand the absence of Landau pole up to some
large value $\Phi_0$. So the restriction is:  $0 < \lambda{(t)} < \infty$
up to the scale $\Lambda$ at which new physics begins. In order to get
$\lambda(t)$ behavior one needs the renormalization group equations
which determine the behavior of $\lambda$, Yukawa coupling constants
of $t$-quark and new quarks and leptons  with Higgs doublet and gauge
coupling constants.  These equations can be found in literature
\cite{5}, \cite{200}.  Let us present here the renormalization group
equation for the running value $\lambda(t)$ in a theory with $N$ generations
of heavy fermions with the degenerate masses \footnote{Similar problem for
$N=1$ was considered in \cite{200}. However, our numerical results differ
since in \cite{200} fourth neutrino was massless, and $M_{top}$ was
considered as a free parameter.}:
\begin{eqnarray}
\nonumber
\frac{d\lambda}{dt} = \frac{3}{2\pi^2}
\lambda^2 + \frac{\lambda}{4\pi^2} [3g^2_t + 3N(g^2_T + g^2_B) + N(g^2_E +
g^2_N)] -\\
\nonumber - \frac{1}{8\pi^2}[3g^4_t + 3N(g^4_T + g^4_B) +
N(g^4_E + g^4_N)] - \\
\nonumber - \frac{3}{16\pi^2} \lambda(3g^2 +
g'^2) + \frac{9}{384\pi^2} g'^4 + \frac{9}{192\pi^2} g^2g'^2 +
+\frac{27}{384\pi^2} g^4\;\;,
\label{2}
\end{eqnarray}
where $g$ and $g'$ are SU(2) and U(1) coupling constants and constants $g_i$
determine the masses of the corresponding fermions by  the
following formula:  $M_i = \frac{g_i(0)\eta}{\sqrt{2}}$, $g_t(0) =
1.035$ corresponding to $M_t = 180$ GeV.  Performing calculations we
add to equation (\ref{2}) the renormgroup equations for $g, g'$,
SU(3) coupling  constant $g_3$ and
$g_i$ and  numerically integrate the system of the coupled
differential equations.  (Details will be published later).
\begin{figure}
\epsfbox{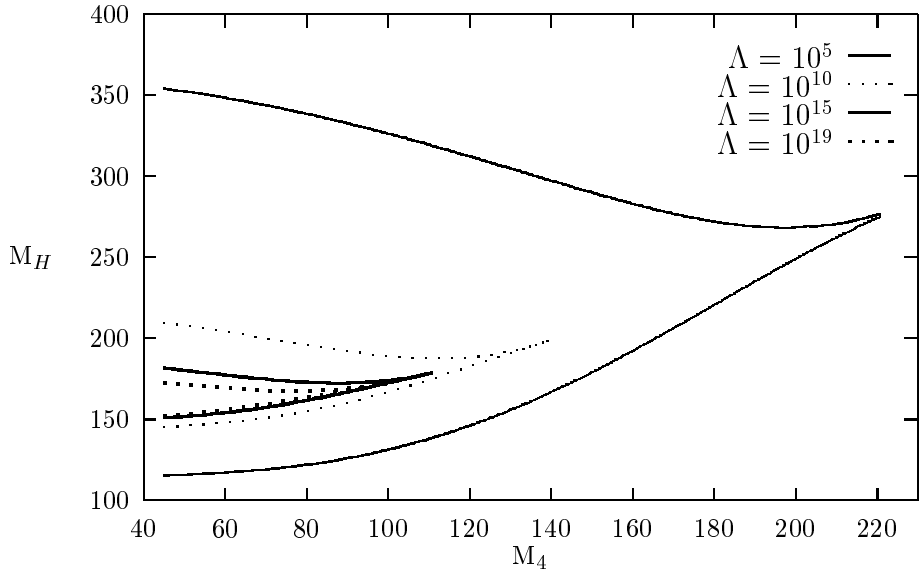}
\caption{ Allowed values of $M_H$ and $M_4$ lie between two curves:
a. solid for $\Lambda = 10^5$ GeV;
b. thin dotted for $\Lambda = 10^{10}$ GeV;
c. thick solid for $\Lambda = 10^{15}$ GeV;
d. thick dotted for $\Lambda = 10^{19}$ GeV. }
\end{figure}
Results of the calculations for the case of one heavy generation are
presented in Fig. 1.

If the value of $M_4$ is small, then we approximately get the
allowed interval
of the values of $M_H$ for $M_t = 180$ GeV
in the Standard  Model, well-known from literature. For all values of
ultraviolet cutoff
$\Lambda$ the low lines which represent $\lambda = 0$ stability
bound go up with growing $M_4$.  The physical reason for such
behavior is clear -- the third term in equation (\ref{2}) becomes
larger and a heavier Higgs is required to get a positive potential
for heavier fermions. The upper lines which represent Landau pole
bound are governed by the first term in (\ref{2}) and are almost
constant for $\Lambda > 10^{10}$ GeV.  However, for $\Lambda = 10^5$
GeV new phenomena occur -- the second term in (\ref{2}) becomes
essential and an upper curve goes down for increasing $M_4$. So
the allowed interval of $M_4$ values shrinks.
\begin{figure}
\epsfysize=7cm
\epsfbox{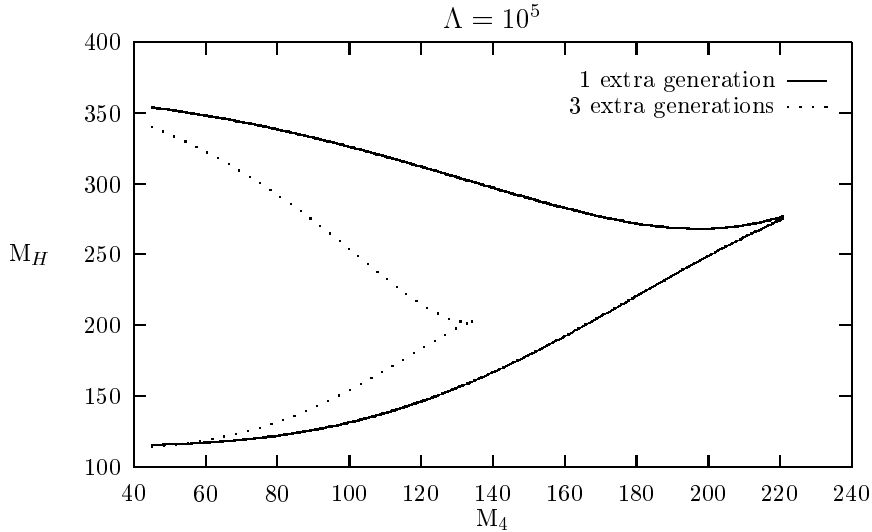}
\caption{ Allowed values of $M_H$ and $M_4$ for 1 and 3 extra
generations for $\Lambda = 10^5$ GeV. }
\end{figure}

We see that a heavy generation  can
exist only if new physics starts below $10^{10}$ GeV. The mass of this
generation should be smaller than 220 GeV while the possible values
of $M_H$ vary between 150 and 300 GeV if $\Lambda = 10^5$ GeV.

The introduction of the additional heavy generations will
restrict the allowed values of $M_{extra}$ and $M_{Higgs}$ (Fig. 2 - Fig. 4 ).
 For
example, for $N=3$ and $\Lambda = 100$ TeV only $M_{extra} < 140$ GeV
is allowed (see Fig. 2).

We should notice that the precise electroweak measurements can hardly allow
3 extra  generations while one extra generation cannot be excluded
\cite{3}, \cite{100}.
\begin{figure}
\epsfbox{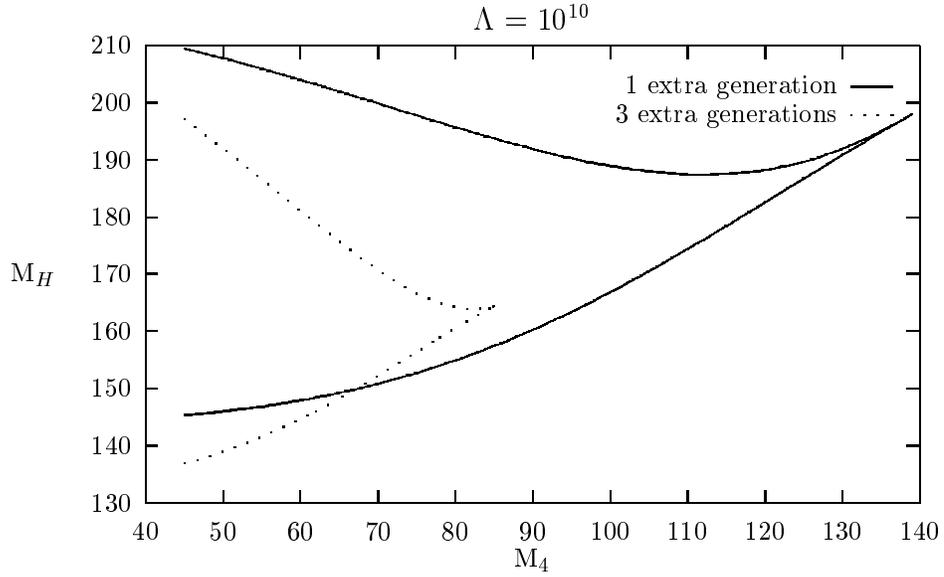}
\caption{ The same as Fig. 2 for $\Lambda = 10^{10}$ GeV. }
\end{figure}
\begin{figure}
\epsfbox{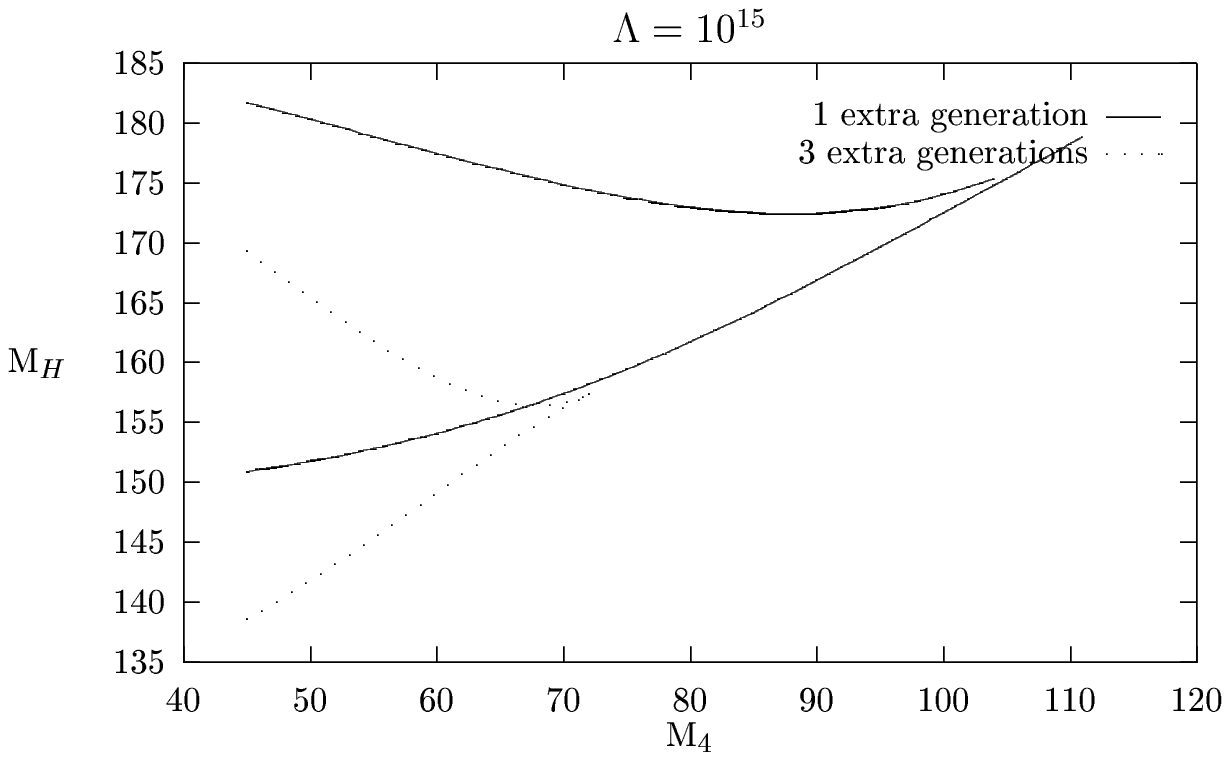}
\caption{ The same as Fig. 2 for $\Lambda = 10^{15}$ GeV. }
\end{figure}
\begin{figure}
\epsfbox{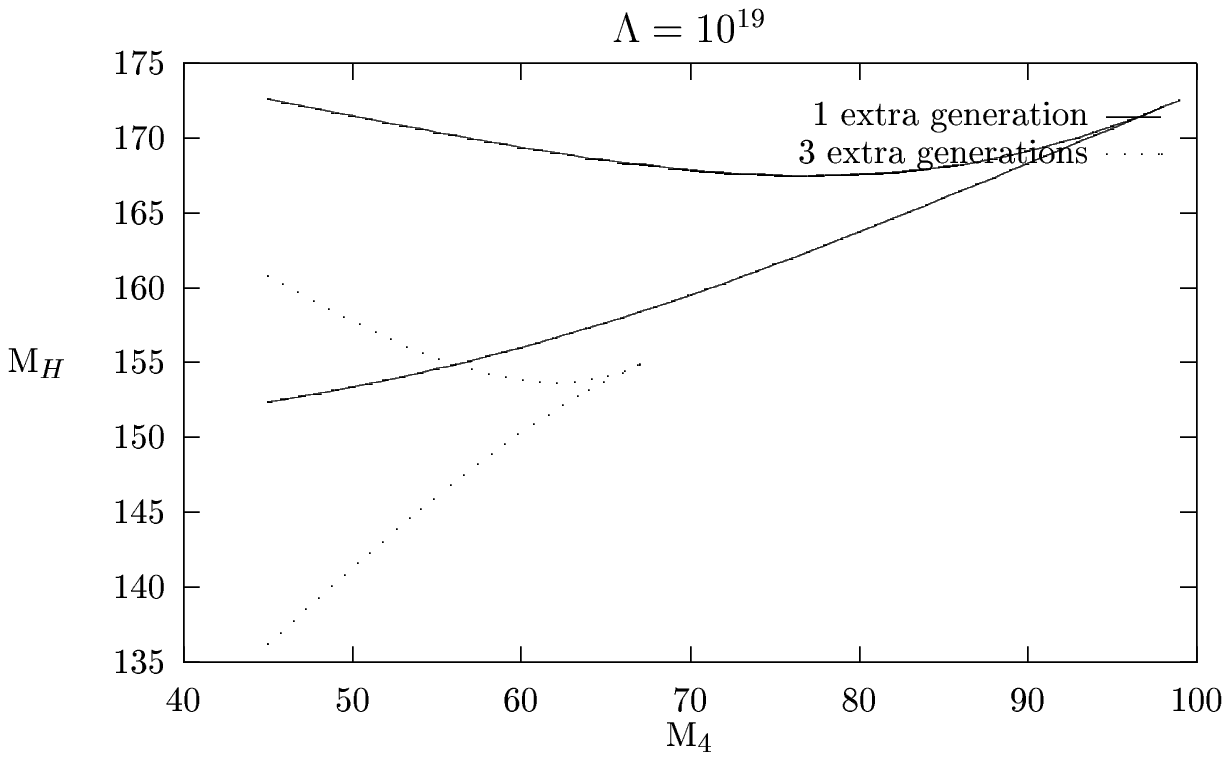}
\caption{ The same as Fig. 2 for $\Lambda = 10^{19}$ GeV. }
\end{figure}
Further predictions of the Higgs boson and extra generation masses are
possible with the help of multiple criticality principle,
which correspond to the degenerate minima of Higgs potential at
$\eta$ and $M_{Pl}$ \cite{6}.
We find that for $M_t = 180$ GeV this principle forbids the existence of
fourth generation, i.e. there are no such values of $M_H$ and $M_4$
that at $\Lambda = 10^{19}$ GeV effective one loop potential has
degenerate minima.

In our analysis we use the one-loop renormalization group improved
potential.  The shift of numbers obtained due to second loop should
not be very large, see e.g.  \cite{301} Altarelli, Isidori, where it is
shown that for the case of the standard model with 3
quark-lepton generations a lower bound on Higgs mass is shifted by
the second loop 10 GeV down.

We are grateful to C. Froggatt, L. B. Okun, and K. A. Ter-Martyrosyan for
useful
discussions, to Dr. Jensen for help in finding ref. \cite{10} and to
L.B.Okun who brought our attention to ref. \cite{200}.
H. B. Nielsen, V.A.Novikov and M. I. Vysotsky were partially supported by
INTAS grant 93-3316, A.V.Novikov, V. A. Novikov and M. I. Vysotsky were
partially
supported by INTAS grant
94-2352 and RFFR grant 93-02-14431.

\end{document}